\def\frac#1#2{{\begingroup#1\endgroup\over#2}}
\documentstyle[paspconf,epsfig]{article} 
\def\epsscale#1{\gdef\eps@scaling{#1}}
\def\plottwo#1#2{\centering \leavevmode
    \epsfxsize=.45\columnwidth \epsfbox{#1} \hfil
    \epsfxsize=.45\columnwidth \epsfbox{#2}}

\begin{document} 

\title{Ly$\alpha$  forest and the total
 absorption cross-section of
galaxies -- an example of  the NTT SUSI Deep Field} 

\author{Srdjan Samurovi\'c\altaffilmark{1}}
\affil{Public Observatory, Gornji Grad 16, 11000 Belgrade, SERBIA}
\author{Milan M.~\'{C}irkovi\'{c}\altaffilmark{2}}
\affil{Dept. of Physics \& Astronomy, 
SUNY at Stony Brook, 
Stony Brook, NY 11794-3800, USA}
\affil{Astronomical Observatory, Volgina 7, 11000 Belgrade, SERBIA}

\altaffiltext{1}{email: srdjanss@afrodita.rcub.bg.ac.yu }
\altaffiltext{2}{email: cirkovic@sbast3.ess.sunysb.edu }
\begin{abstract}
By extrapolating the accumulated low-redshift data on the absorption radius of galaxies
and its luminosity scaling, it is possible to predict the total absorption cross-section
of the gas associated with collapsed structures in the universe at any given epoch. This 
prediction can be verified observationally through comparison with the well-known spatial
distribution of the QSO absorption systems. In this way, it is shown that HDF, NTT SUSI
Deep Field and other such data give further evidence for the plausibility of origin of 
the significant fraction of the Ly$\alpha$ forest in haloes of normal galaxies.
\end{abstract}
\section{Introduction}

 Deep images like the {\it Hubble Deep Field} (HDF), 
fields of the QSO BR1202-0725 (Giallongo et al. 1998) and NTT SUSI Deep
 Field (NDF) (Arnouts et al. 1998), offer an unprecedented opportunity to
  study global properties of galaxies at all epochs. One of the crucial 
  such property with relevance to several fields of astrophysical research 
  is the absorption cross-section of normal luminous galaxies. From 
  coincidence studies of metal and Ly$\alpha$ absorption lines seen 
  in QSO spectra, it was recently established that significant fraction 
  (possibly all) of the low-redshift ($z<1$) absorbers are associated
   with galaxies, with known luminosity scaling (Steidel, Dickinson \& 
   Persson 1994; Lanzetta et al. 1995; Chen et al. 1998). It is only natural 
   to ask whether the same situation persists when we look to higher 
   redshifts and what is appropriate total absorption cross section of
    the universe due to galactic haloes.  

\section{Comparison of deep fields}

A crude way to estimate the total absorption cross-section and to predict
 the number of narrow Ly$\alpha$ absorption lines to be observed in spectrum
 of a source located at arbitrary $z$ is to use the galaxy surface densities
 obtained from very deep images and extrapolate the results of low-redshift
 analysis up to redshifts of $z \sim 4$, at which redshifts are easily
 accessible today through photometric techniques (e.g. Lanzetta,
 Yahil \& Fernandez-Soto 1996). 

In the ``zeroth'' approximation, we have assumed identical sizes of
 absorbing galaxies, and compared the predictions for the total
 number of absorption lines
 obtained in this manner in the HDF and in the field of the QSO BR1202-0725, 
as well as both of them with the empirical data on the spatial
 distribution of the Ly$\alpha$ forest. These results are 
shown in the Fig. 1. Calculations were done within context of Einstein-de 
Sitter universe, and are independent on the value of $H_0$. The covering 
factor here and elsewhere is assumed to be unity (\'Cirkovi\'c et al. 1997).

\begin{figure}
\epsscale{0.10}
{\plottwo{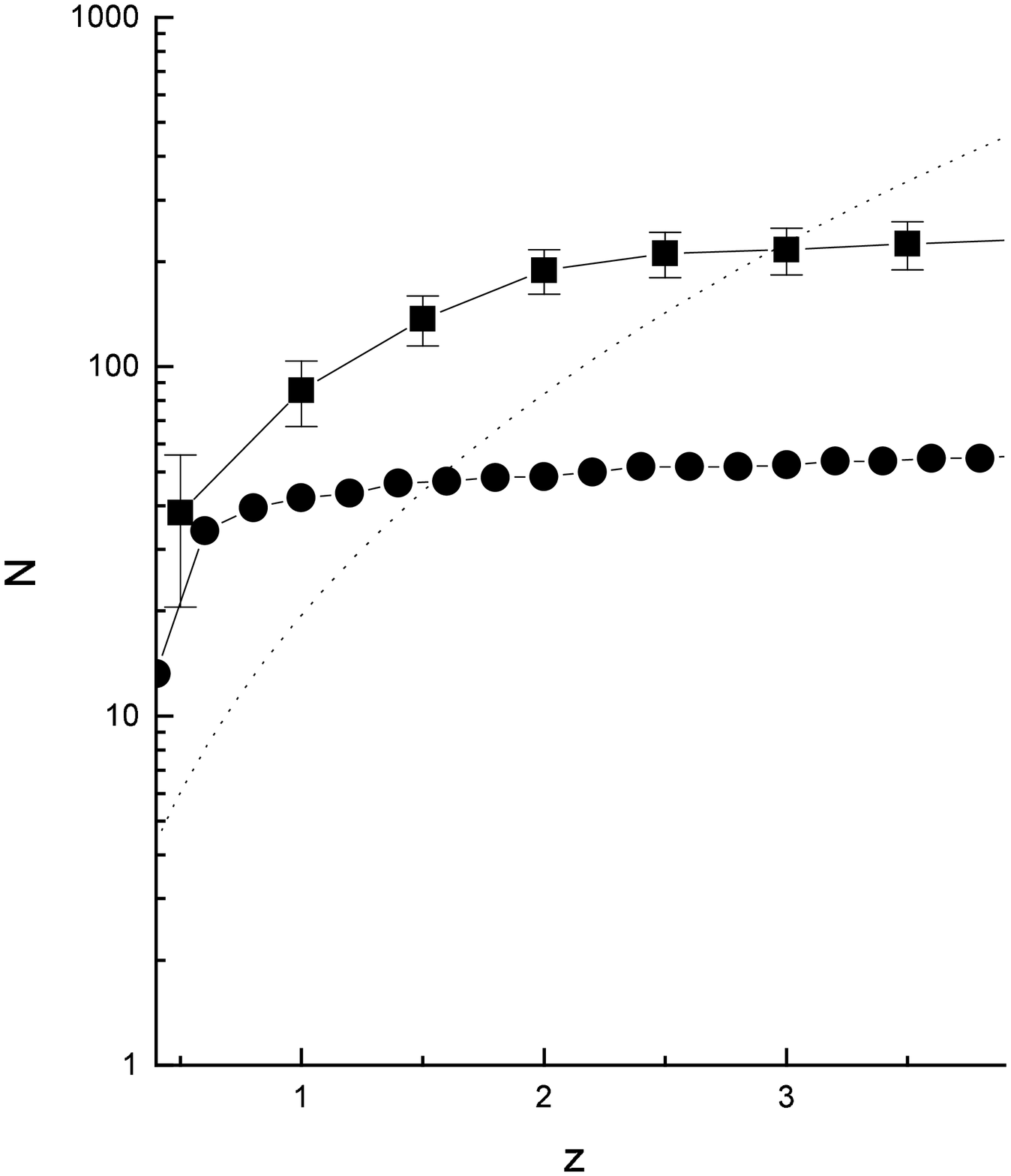}{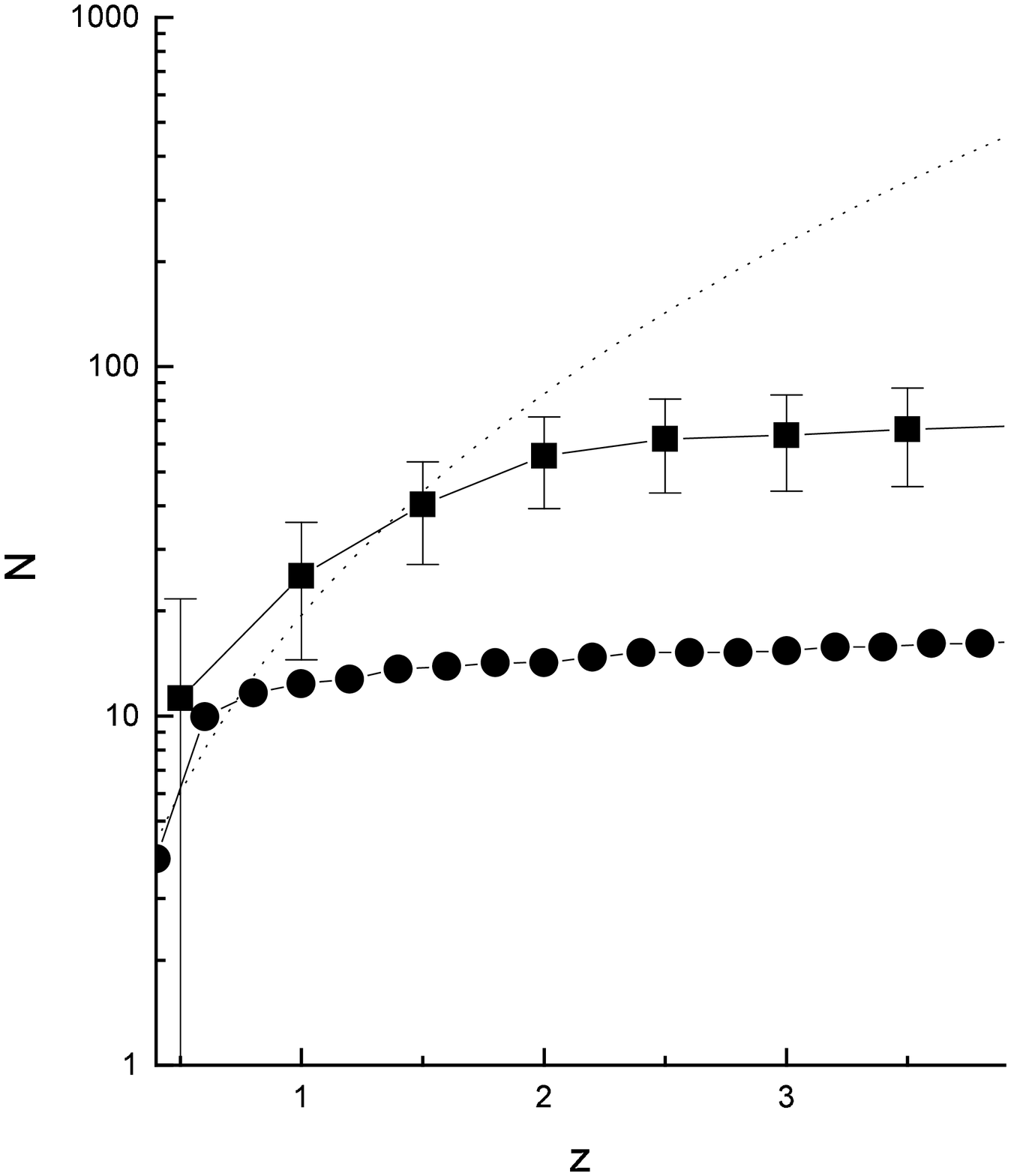}}
\caption{Left: The number of predicted absorption
 lines $ N$ with 1${\sigma}$ error bars (for the 
HDF data) 
as a function of redshift, without luminosity scaling, in
 comparison with
 the empirical data derived from absorption statistics (dotted curve). 
Points in the field of BR1202-0725 are shown as filled circles.
Right: The same as in the left panel, with 
luminosity scaling according to the data by Chen et al. (1998).}
\end{figure}

Next natural step is to use recently established luminosity scaling of
 the absorbing material associated with galaxies with galactic B-band
 luminosity. Convolution of the local luminosity function (Schechter
 1976; Willmer 1997) with available photometric 
surface densities acts, as expected, to reduce the predicted number
 of absorption lines and brings it into better agreement with the empirical
 data, as shown in the right panel of Figure 1. In the simplest (Einstein-de Sitter) case,
 the total number of absorption lines seen out to redshift $z$ is given as
$$N=\frac{\kappa \pi}{4} \frac{R_0^2 H_0^2}{c^2} \left(
\frac{10800}{\pi} \right)^2 \int_0^z n(z) \frac{(1+z)^3}{(\sqrt{1+z}-1)^2}
 \times 
$$

 $$
\;\;\;\;\;\;\;\times \left[
\frac{\phi_\ast}{L_\ast} \int_{L_{\rm min}}^{L_{\rm max}} \left(\frac{L}
{L_{\ast}} \right)^{2\alpha -\gamma}
 \exp \left( -\frac{L}{L_\ast} \right) dL \right] dz,\eqno(1)$$ 
where $\gamma$ and $\phi_\ast$ are the parameters of the luminosity 
function, $n(z)$ is the galaxy surface density at $z$ in (arc min)$^{-2}$, 
and $\alpha$ is the index of Holmberg-type scaling of the absorption radius
 (Chen et al. 1998). Only luminosities corresponding to magnitudes
 $-21.5 < M < -14$ are considered (Willmer 1997), and standard Schechter 
luminosity is taken to correspond to absolute magnitude $M_\ast =-19.1$.

The agreement of predicted and observed spatial distribution of absorption 
lines at low and intermediate redshift is obvious, especially in the case
 of the HDF. Unfortunately, the errors in surface densities for the field of 
BR1202-0725 are still not available, but it is probable that  
agreement at lower redshifts is achieved here too. Empirical power law for the 
spatial distribution of Ly$\alpha$ forest was used with the values 
of Kim et al. (1997). Our results are in agreement with those of 
Fernandez-Soto et al. (1997), although we do not find a single power 
law capable of reproducing data points at low $z$ with any statistical
 significance. 
We notice that these results suggest the existence of a critical redshift
 $z_c$ above which it is impossible to explain observed absorption with the
 material exclusively associated with galaxies. This is expected,
 in view of several other circumstantial arguments, such as behavior of
 the Ly$\alpha$ forest autocorrelation function (e.g. Cristiani et al. 1997),
 as well as the results of numerical simulations of the structure formation.
 The theoretical significance of that result as one further discriminator
 between various gasdynamical histories will be discussed in a subsequent
 work. 
 
In order to investigate whether the ground-based deep fields like that of
 BR1202-0725 are representative of the high-redshift galaxy population, 
we have performed several other tests, including analysis of the angular
 size distribution of sources in the NDF. Due to spatial limitations, we do 
not present results here, but emphasize the strong impression that,
 within uncertainties due to large pixel size, the field is fairly 
representative sample of galaxy populations, at least up to redshifts 
of $\sim 3$, which are interesting from our current point of view. 

Other issues to be aware of are dependence of angular size on the cosmological
 model (models with $\Lambda$-term are interesting in this respect), 
and changes in the comoving flux in the selected band due to effects of
 secular galactic evolution. At this point, the problem of the total 
absorption cross-section of galaxies comes again into close contact with the
 general problem of star-formation history of the universe. 
Further problems to be addressed in the subsequent work are various effects 
connected with non-conservation of the phase-space density of galaxies. 
This arises for fundamentally two reasons: late ($z<3$) galaxy formation,
 predicted in many models of structure formation, and mergers of galaxies. 
Both effects have been only slightly discussed in a quantitative way so far.

Work currently in progress includes analysis of NDF photometric redshifts, 
and their incorporation in the general pattern, as well as more precise and
 accurate analysis of errors and uncertainties, apart from the points 
mentioned above. Further significance stems from the prospects of incoming 
deep images, like Southern Deep Field, to be taken in very near future.
 It is our modest hope that these results will, together with all others  
resulting from the current revolution in observational techniques,
 contribute to our understanding of the structure and evolution of the deep 
universe. 

\begin{figure}

\plottwo {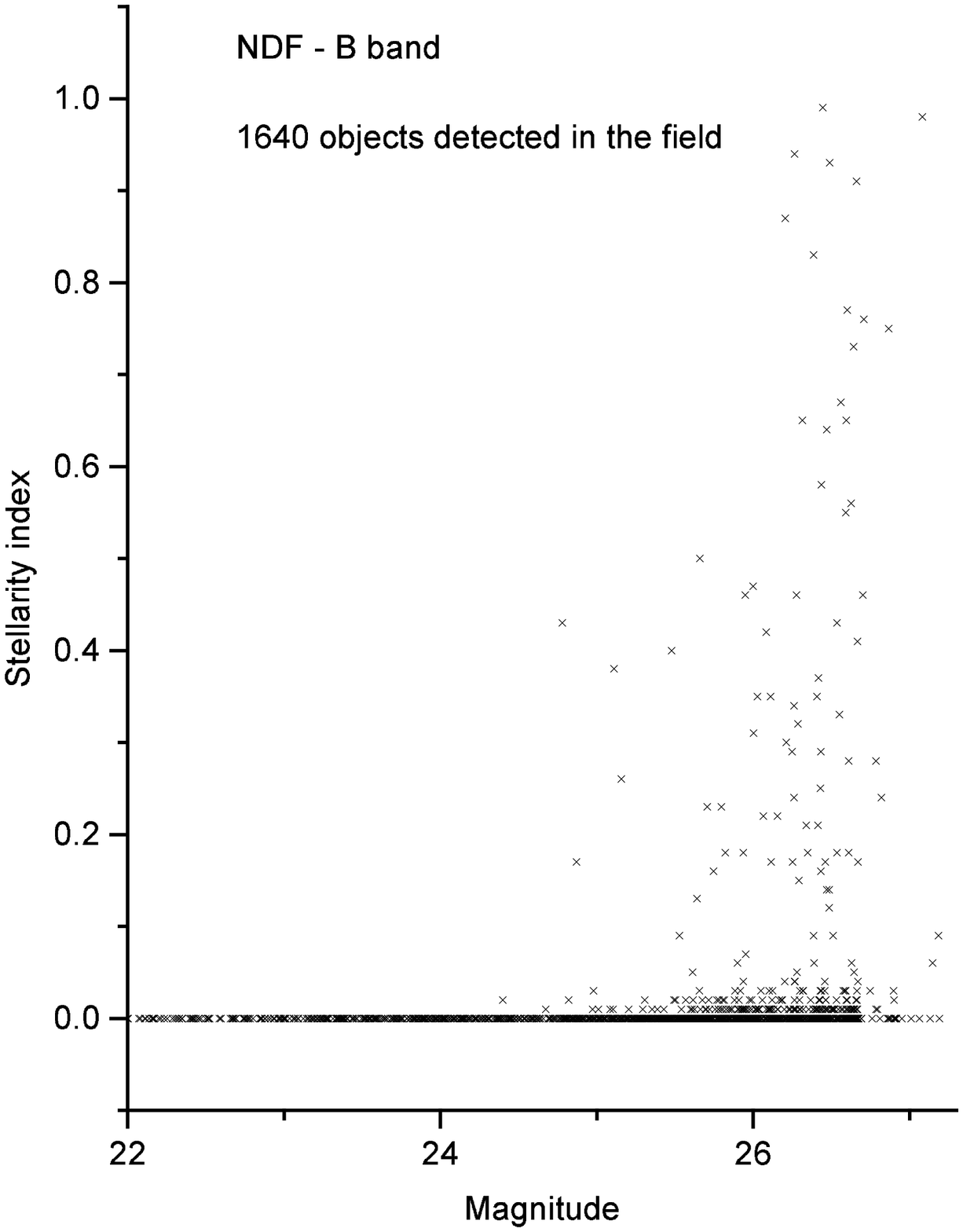}{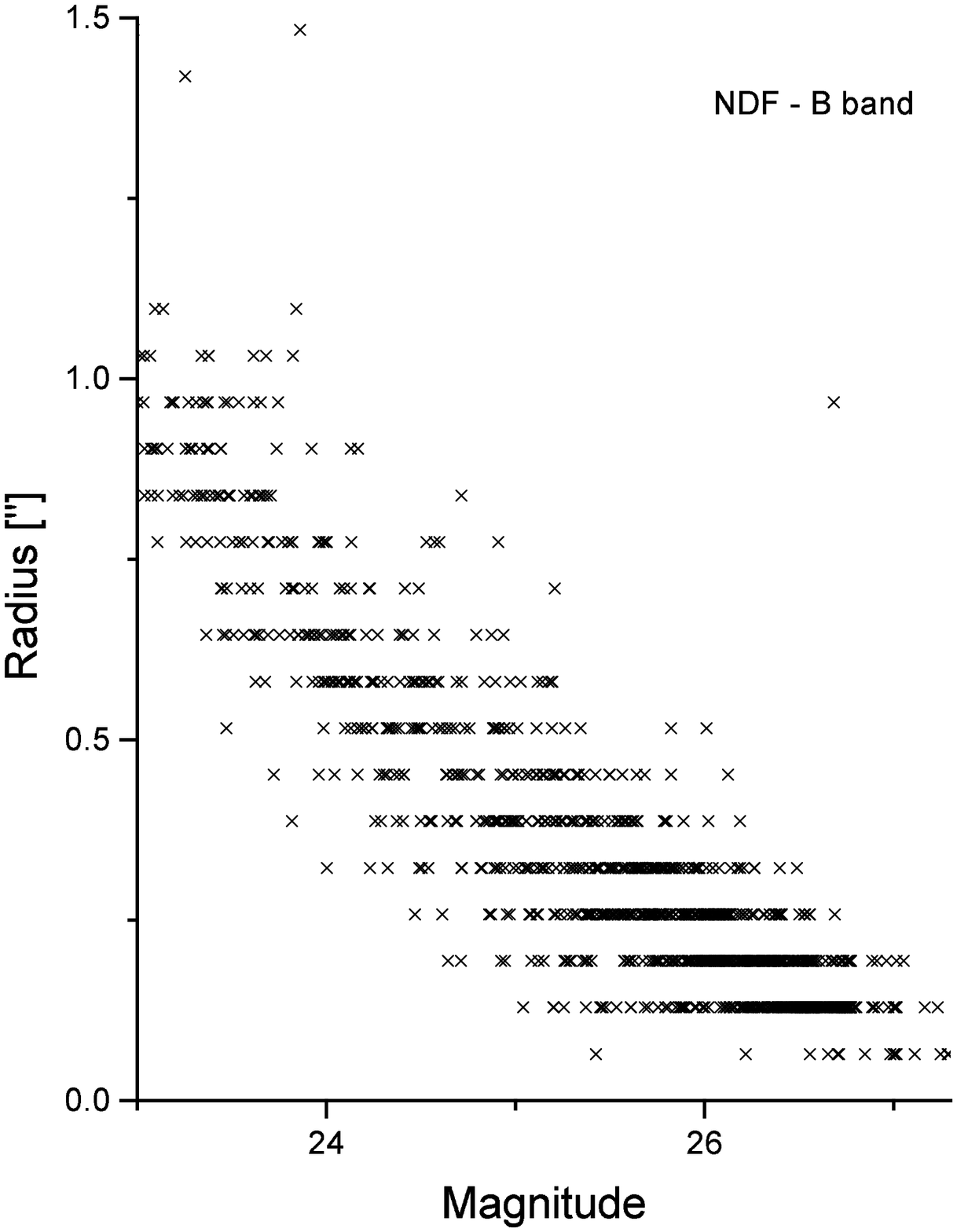}

\caption{Left: Stellarity index as a function of magnitude.
Right: Angular radius as a function of magnitude for the B band.}

\end{figure}

\begin{figure}
\plottwo{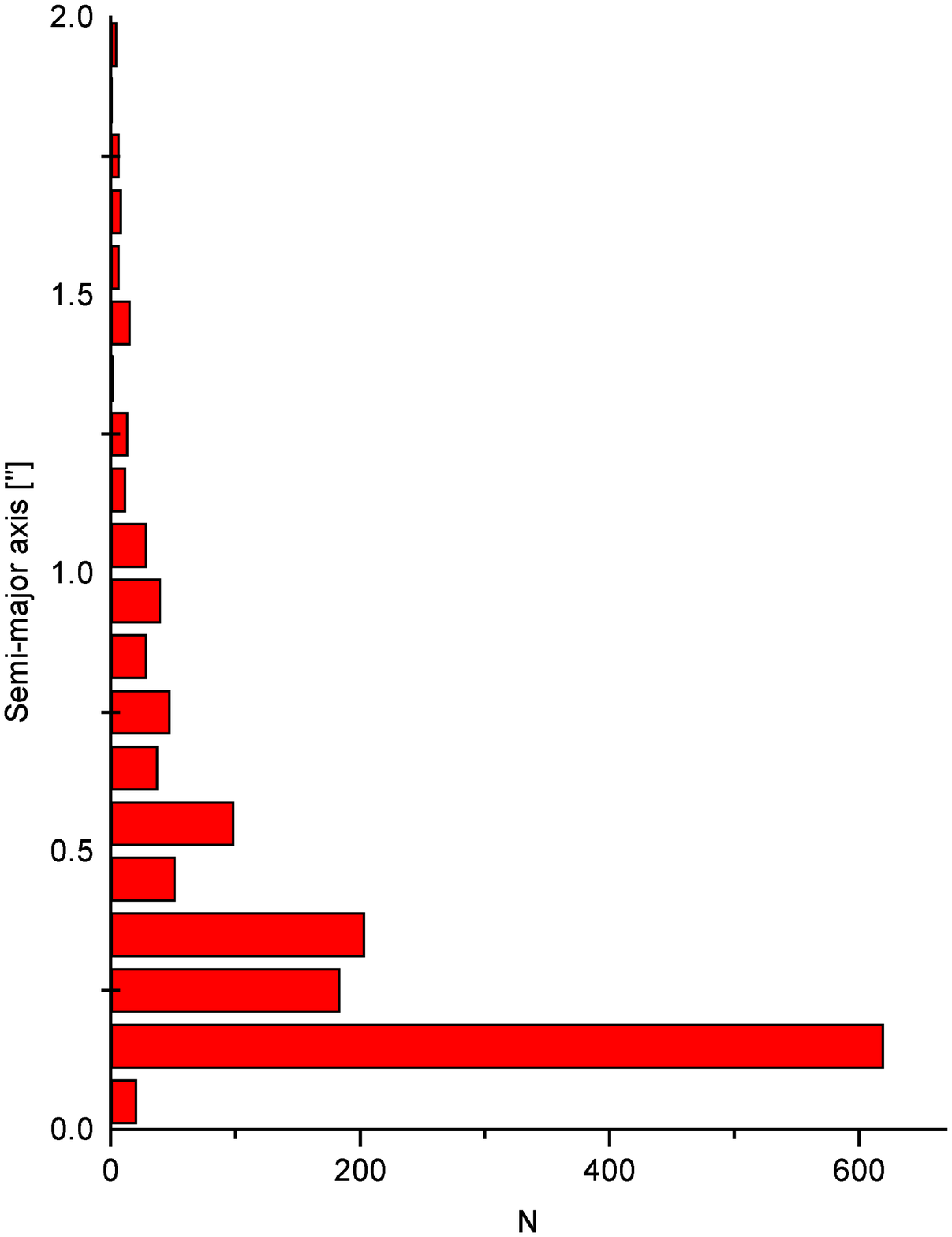}{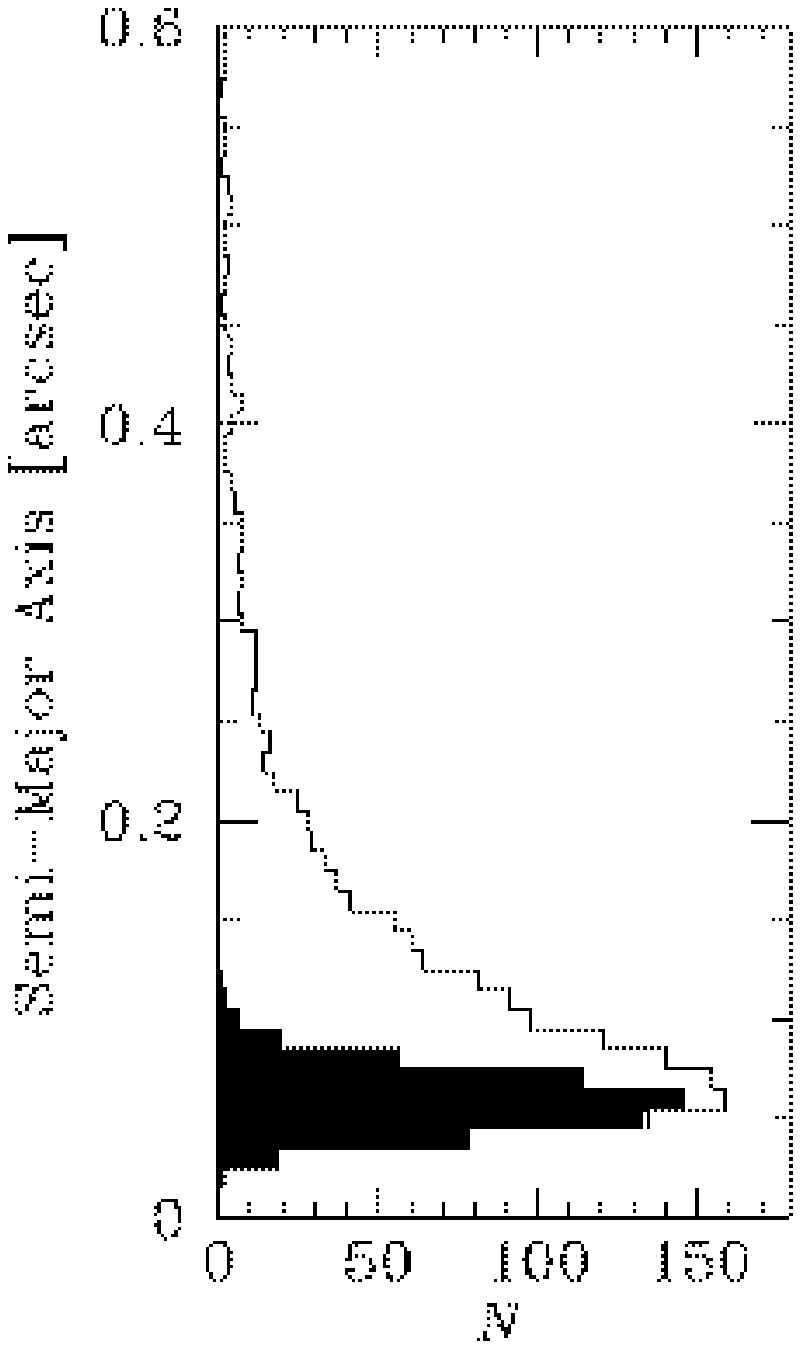}

\caption{Left: Histogram of semi-major axes of NDF
 galaxies up to AB $ \sim 27$. Right:  The same as in the left panel
  for HDF (from Yahil et al.~1998) with correction for incompleteness at AB$>28$ (shaded region).}
 \end{figure}

\section {Technical remarks and some results}

We used the SExtractor program (Bertin and Arnouts 1996) in 
order to analyze the NTT SUSI Deep Field images. 
In the Figure 2 (left panel) we
 present the plot of the ``stellarity index''  that
 ranges between 0.0 (galaxy) and 1.0 (star) for the B band; we included
all extracted objects regardless on the {\tt S/G} ratio and {\tt flag}
parameter.
We also present some preliminary results concerning angular diameter of the
 galaxies in the NTT SUSI Deep Field for the B band (Figure 2, right). We note that
 the ``holes'' in the distribution are artifacts due to 
the rather large size of the pixel (1 pixel
 corresponds to 0.129 arcsec). This result can be compared to other 
estimates (e.g. Shanks et al. 1998). Further analysis is to be performed 
after obtaining the redshift estimates for the galaxies in the field 
in order to test the effects of non-zero cosmological constant, $\Lambda$.
Figure 3 (left panel) shows the histogram of the number of galaxies according to
their angular radius. A comparison with the corresponding data obtained 
from the HDF analysis (Yahil, Lanzetta and Fernandez-Soto 1998), presented
 in Figure 3 (right panel), shows obvious similarities.  
We exclude the galaxies with {\tt S/G} $\ge$ 0.5 (cf.
 Arnouts et al. 1998) and {\tt flag} $\ge$ 10;
  this criterion is quite conservative, as can easily be seen from the Figure 2 (left panel) 
  and in comparison to the quoted references.

\acknowledgements

We thank Dr.~Sandro D'Odorico for sending us  the draft
 version of the NTT SUSI Deep Field paper. We acknowledge ESO for providing
NTT SUSI Deep Field FITS files via WWW.

\end{document}